\documentclass[preprint]{elsarticle}[12pt]
\usepackage{epsfig}
\usepackage{supertabular}

\begin{document}

\title{Information Theory and Population Genetics}

\date{July 2, 2011}
\author{Reginald D. Smith\corref{cor1}\fnref{fn1}}
\ead{rsmith@bouchet-franklin.org}
\address{PO Box 10051, Rochester, NY 14610 USA}
\cortext[cor1]{Corresponding author}
\fntext[fn1]{Phone: +1 585-244-4979}

\begin{abstract}
The key findings of classical population genetics are derived using a framework based on information theory using the entropies of the allele frequency distribution as a basis. The common results for drift, mutation, selection, and gene flow will be rewritten both in terms of information theoretic measurements and used to draw the classic conclusions for balance conditions and common features of one locus dynamics. Linkage disequilibrium will also be discussed including the relationship between $I$ and $r^2$.
\end{abstract}


\maketitle
\section{Introduction}
Population genetics and information theory both began to emerge in the first half of the 20th century. Population genetics, animated by the ongoing debate about the relationship between the theory of evolution, driven by natural selection, and the laws of Mendelian inheritance, became one of the foundations of modern biology and enabled biologists to show how the frequency of inherited alleles as well as genotype frequencies in a population can be affected by the various processes such as mutation, selection, and genetic drift \cite{popgen1,popgen2,popgen3,popgen4,popgen5}. With the rise of the neutral theory of evolution \cite{kimura,kimura2} and the genomics revolution, it has helped supplement the insights gained from genetic data and been used to explain phenomena such as the ratio and rates of synonymous and non-synonymous substitutions and how this can be used both as a molecular clock between species or to identify positively or negatively selected genes \cite{dNdS}, coalescent theory which addresses the distance between populations separated by time but linked due to a recent common ancestor \cite{coalescent}, and gene flow amongst genetically modified and wild organisms \cite{GMO}.

Information theory, though born a couple of decades after most of the initial insights of population genetics, has had an impact just as far reaching and important. Developed by the legendary Claude Shannon in the 1940s at Bell Labs \cite{shannon}, information theory enabled the communications revolution, the Internet, and revolutionized views of entropy and information allowing information and information transfer rates to be successfully quantified. In a tribute to its utility and expansive scope, information theory was later adopted by other disciplines to understand related or completely unrelated phenomena \cite{bandwagon}. Some of the best known examples are the papers by Edwin Jaynes showing that much equilibrium statistical mechanics can be derived using only information theory and assumptions of maximum entropy \cite{jaynes1,jaynes2,jaynes3}. This has led to the rise of the Maximum Entropy (MaxEnt) school of inquiry in statistics and the expanded use of information theory across a wide variety of the natural sciences. Interestingly enough, in his Ph.D thesis written at Cold Spring Harbor Laboratory, Shannon tackled the ideas of population genetics \cite{shannonpopgen}. A concise and fascinating summary of his work is given by James Crow \cite{crowshannon}.

In this paper, it will be shown that there are deep links between quantitative population genetics and information theory. This will not be an abstract treatment with only a passing reference to biologically meaningful and important quantities. It also is not an attempt to claim that the underlying mechanisms of evolution are based on ``information'', vaguely defined, instead of well-understood and recognized biological forces. Rather, this will show that the key valuable results of population genetics can be understood by seeing that the evolution of allele frequencies in a population can be interpreted as a biological process whose mechanisms have exactly corresponding information theoretic measures and that the techniques of information theory can shine new light on what these biological processes mean in aggregate as well as simplifying theoretical analysis of some evolutionary processes.

\section{Preliminary Concepts}

There have been prior investigations of population genetics borrowing tools from information theory. First, for years some of the most popular metrics for measuring biological diversity have been borrowed from information theory \cite{diversity1,diversity2,diversity3}. It is difficult to say exactly when the subject of a role for entropy in population genetics was first tentatively raised. Moran did investigate the entropy of general Markov processes \cite{moran} in a now almost forgotten paper written after his famous papers on birth-death population genetics models. Watterson approached the subject again a year later in his 1962 paper \cite{watterson} on diffusion theories in population genetics. Towards the end of the paper, he calculated the entropy of the allele frequency distribution as a possible measure of the time for a population to completely lose one allele, decaying towards homozygosity. As will be shown later in this paper, he was very correct as entropy does directly determine the decay time of a population's heterozygosity by genetic drift. Also, other works seek to explain or derive aspects of population genetics using techniques involving Fisher information \cite{popinfo1} or general methods of computation for nonlinear dynamical systems \cite{popinfo2}.

This treatment will be different, however, in showing that all evolutionary forces can be consistently represented by information theoretic measurements in a comprehensive theory. In this paper, the focus will be on the two allele single locus model. Allele frequencies, $p$ and $q$, will be represented as objective probabilities which will reflect the presence of the alleles amongst all loci in the population. The connections between population genetics and information theory will be made using several information-based quantities which will be defined in this section.

First, and most famous in information theory, is the concept of Shannon entropy. For a random variable distribution with $n$ different states with probabilities $P(i)$ where $\sum_{i=1}^n P(i) = 1$, the entropy, $S$, of the distribution is defined by

\begin{equation}
S = -\sum_{i=1}^n P(i)\log P(i) 
\label{entropy}
\end{equation}

The value of $S$ always ranges between a minimum of 0 for the trivial distribution where an event occurs with a probability 1 and a maximum of $S = \log n$ for the uniform distribution across all $n$ states. This simple definition of entropy belies the fact that entropy has several orders depending on the degrees of freedom defined in the distribution. The lowest order, zeroth order entropy, or $S_0$, is simply represented by 

\begin{equation}
S_0 = \log n
\label{entropy0}
\end{equation}

and depends on only the number of possible states in the distribution. For the two allele model, $S_0 = \log 2$. The first order entropy, which will be referred to as $S$ without a subscript, is the traditional definition given in equation \ref{entropy} and its maximum value is the value of $S_0$. The second order entropy, $S_2$, the last relevant order for two allele models, is given by

\begin{equation}
S_2 = -\sum_{i=1}^n \sum_{j=1}^n P(i,j)\log P(i,j)
\label{entropy2}
\end{equation}

The quantity $S_2$ is often described as the joint entropy and is based on the joint probability of states $i$ and $j$. In the paper, it will also be referred to as $S$ over two variables, e.g. $S(x,y)$. Similar to the relationship between $S_0$ and $S$, $S_2 \leq S(i) + S(j)= 2S$.

In addition to the measures of entropy, there will be two other useful quantities, the Kullback-Leibler divergence, $D$, and mutual information, $I$. The Kullback-Leibler divergence is a quantity which measures the difference between two probability distributions. For two distributions, $f$ and $g$, the  $D$ from $f$ to $g$ is defined by

\begin{equation}
D(f,g) =\sum_{i=1}^n f(i) \log \frac{f(i)}{g(i)}
\label{kd}
\end{equation}

where $D \geq 0$. One important aspect of the divergence to note is it is not a distance metric since $D$ is not symmetric with respect to the distances between the distributions and $D(f,g) \neq D(g,f)$. Another way to express $D$ is

\begin{equation}
D(f,g) = S_X(f,g) - S(f)
\label{kd2}
\end{equation}

where $S_X$ is the cross entropy represented by $S_X = -\sum_{i=1}^n f(i) \log g(i)$

The Kullback-Leibler divergence will be integral to our discussion of genetic drift.

The mutual information, $I$, between two random variables, $i$ and $j$ is a representation of the entropy from one variable that can be derived given the entropy of another in a distribution. Shannon first used it to measure the capacity of a channel by seeing how much the output of message at the receiver could be determined by the input. Formally, $I$ is given by

\begin{equation}
I = \sum_{i=1}^n \sum_{j=1}^n P(i,j)\log \frac{P(i,j)}{P(i)P(j)}
\label{MI}
\end{equation}

The mutual information also has an alternate formulation

\begin{equation}
I = S(i) + S(j) - S(i,j) = S(i) - S(i|j) =  S(j) - S(j|i)
\label{MI2}
\end{equation}

Mutual information will be used to represent the effects of selection and non-random mating in populations. These quantities will be shown for reference in table \ref{keyinfo} along with their relative significance.

\begin{table}
{\small
\begin{tabular}{|c|c|p{5cm}|}
\hline
Measure&Equation&Significance\\
\hline 
$S_0$&$\log n$&Maximum entropy of an $n$ allele model\\
\hline 
$S_1$&$-\sum_{i=1}^n P(i)\log P(i)$&Entropy of allele frequencies; key measure of change in allele frequencies over time\\
\hline 
$S_2$&$-\sum_{i=1}^n \sum_{j=1}^n P(i,j)\log P(i,j)$&Entropy of loci based on allele pair frequencies; key measure of changes in genotype\\
\hline 
Kullback-Leibler Divergence&$\sum_{i=1}^n f(i) \log \frac{f(i)}{g(i)}$&Used to model genetic drift\\
\hline 
Mutual Information&$\sum_{i=1}^n \sum_{j=1}^n P(i,j)\log \frac{P(i,j)}{P(i)P(j)}$&Used to model selection and non-random mating\\
\hline 
\end{tabular}
\caption{Key measures of information theory and their significance to the evolution of allele frequencies and genotypes in a population.}
\label{keyinfo}
}
\end{table}

One assumption which will be effectively used throughout the paper is the assumption that the allele entropy is extensive. Therefore, we are able to add the cumulative entropy changing effects to come up with the net change in allele entropy each generation.

\section{The basic model and Hardy-Weinberg Equilibrium}

The basic model for the evolution of allele frequencies over time will be given by measuring the first order entropy change between generations. Given discrete generations, we can utilize the techniques of difference equations \cite{diffeq}. This entropy for alleles $p$ and $q$ at the current time is $S_t$ and the change is represented through a difference equation relation

\begin{equation}
S_t = S_{t-1} + \Delta S
\label{basicchange}
\end{equation}

or

\begin{equation}
\Delta S = S_t - S_{t-1}
\label{basicchange2}
\end{equation}

The change in entropy $\Delta S$ is caused by the accumulated effects of evolutionary forces acting on the population. Hardy-Weinberg equilibrium (HWE) is the basic steady state assumption of the of genotype frequencies amongst populations not undergoing any sort of evolutionary selection or non-random mating to force genotype proportions to differ from those expected from random mating. Given that HWE is a statement of the frequencies of genotypes given allele frequencies, it is obvious that both the first order (allele) and second order (genotype) entropies will need to be used. 

In the most trivial case, $\Delta S=0$. In the presence of random mating (zero mutual information), this equality dictates a condition of maximum equilibrium. The distribution of genotypes given by $S(p,q)$ at maximum equilibrium was first expounded in a paper by Wang, Yuan, and Guo et. al. \cite{HWMaxEnt} in which they use Lagrange multiplier techniques to show that the distribution at maximum equilibrium given allele frequencies $p$ and $q$ is the Hardy-Weinberg equilibrium distribution for genotypes: $(p,p) = p^2; (p,q) = 2pq; (q,q) = q^2$. This was further developed by Zhang \& Zhang \cite{HWMaxEnt2} who expand the analysis to limited cases of multiple alleles. Unfortunately, both papers are only available in simplified Chinese at this time but the mathematical portion of the first is shown in Appendix C. Their result can also be seen from the corollary of zero mutual information ($I=0$) at maximum entropy. It is easy to see then that the term in the logarithm for equation \ref{MI} must equal 1 for all terms where 

\begin{eqnarray}
P(p,p)=p^2\nonumber \\ 
P(p,q)=pq\nonumber \\ 
P(q,p)=pq\nonumber \\ 
P(q,q)=q^2 \nonumber\\ 
\end{eqnarray}

\section{Genetic Drift and Kullback-Leibler Divergence}

The first evolutionary force we will model from an information theoretic perspective is genetic drift. Genetic drift is the tendency for allele frequencies to be affected, to the point of fixation for one allele, due to the statistical effect of sampling errors amongst the survival and reproduction in populations which leads to deviations from Hardy-Weinberg equilibrium and its assumption of stable allele frequencies. In effect, genetic drift is caused by deviations in the subsequent generation's allele frequency caused by stochastic processes. Genetic drift is very sensitive to the size of the population and usually only has significant effects on the order of $2N_e$ generations where $N_e$ is the effective population size.

Genetic drift, while being a completely stochastic effect, in a two allele model has the eventual result over a long time span of fixing one allele and eliminating the other. The fixed allele is completely random though the probability of fixation is equal to the frequency of the allele. This random drift, contrary to most diffusion in physical processes which increase entropy, reduces the overall entropy of the allele frequency distribution until a steady state is reached where $S$ = 0. 

The theory of large deviations \cite{largedev1,largedev2} is a branch of probability theory which describes the probability of deviation of an empirical distribution from its expected theoretical distribution. In the theory of large deviations, the entropy function of the size of a deviation from an expected distribution can usually be represented by the Kullback-Leibler divergence. The divergence can also be connected to a probability of deviation $P$ from the mean value by a formula due to Cr\'{a}mer:

\begin{equation}
\lim_{N\to\infty}-\frac{1}{N} \ln P = D
\label{Cramer}
\end{equation}

where $N$ is the number of trials or particles in the system. In the paper referenced earlier by Watterson \cite{watterson}, he determines that the average time for a population with allele frequency $p$ to decay to homozygosity is roughly equal to the entropy. Here we will approximate using only the entropy and excluding the $-4N/N_e$ term to correct for the ratio of effective to actual population in the population and assume the real and effective populations are equivalent and there is no mutation. Therefore the continuous exponential decay of probability can be represented as

\begin{equation}
p = p_0e^{-\frac{t}{S}}
\end{equation}

Normalizing in terms of generations, the probability for the population to decay to homozygosity in one generation is

\begin{equation}
P = e^{-NS}
\label{Einstein}
\end{equation}

Remarkably, if considering $S_0=0$ and rewriting in terms of $\Delta S$, this expression  is the Einstein fluctuation formula, with the ideal constant $R$ instead set as 1. In addition, assuming the limit approximation is valid and for $N$ in equation \ref{Cramer} being $N=2N_e$ we can write the divergence as

\begin{equation}
D = \frac{1}{2N_e}S
\end{equation}

Finally, solving for $P$ in equation \ref{Cramer} and setting it equal to equation \ref{Einstein} as the Einstein fluctuation formula we have $e^{-ND}=e^{-N\Delta S}$ so $\Delta S=D$. Since $D$ is always positive and the net effect of genetic drift is to reduce entropy, $\Delta S$ is changed by subtracting $D$. The final expression for the entropy change by the divergence is 

\begin{equation}
\Delta S = -\frac{1}{2N_e}S
\label{driftequation2}
\end{equation}

Given that $q = 1- p$ we can represent $S$ in one variable. One of the key discoveries of this paper is that key approximations from population genetics can be derived when a linear approximation of $S$ is taken. Using the famous Mercator approximation of $\log x$ around 1, we can make the approximation that $\log x \approx x - 1$. Therefore, the entropy can be shown as below

\begin{equation}
S = -p \log p - (1-p) \log (1-p) \approx 2p(1-p) = 2pq
\label{approximate}
\end{equation}

This shows that in the linear approximation, entropy is approximately the same magnitude as the heterozygosity frequency, $h$, in the population. Why is this important? Classical population genetics used various scale and linear approximations to deal with the balance equations since advanced nonlinear analysis techniques were then not available. The derivations shown below will take advantage of this showing that these same approximations can be shown to be a linear, limiting case of a more general treatment based on entropy.

Equation \ref{driftequation2} thus becomes

\begin{equation}
\Delta h = -\frac{1}{2N_e}h
\label{heteroequation}
\end{equation}

The form in both equations \ref{driftequation2} and \ref{heteroequation} is the difference equation form for compound growth and the solution for $h_t$ works out to be

\begin{equation}
h_t = h_0\bigg(1-\frac{1}{2N_e}\bigg)^t
\label{heteroequation2}
\end{equation}

with a continuous time expression 

\begin{equation}
h_t = h_0e^{\frac{-t}{2N_e}}
\label{halflife}
\end{equation}

Both of these expressions give a half-life of heterozygosity at $t_{1/2}=2N_e \ln 2$. All of these results completely agree with the calculated decay of heterozygosity and genetic diversity that drift causes. As the heterozygosity approximation will often successfully be used in the paper, a few caveats are needed. The entropy and heterozygosity are best used to look at similar behavior during the evolution of the population. This can lead to valid theoretical insight, however, entropy should not be used as a numerical proxy for the exact value of heterozygosity as these can differ in value while showing the same overall behavior. Also, the entropy as heterozygosity approximation is only valid when there is random mating and only the effects of drift, mutation, or migration are impacting. As will be explained later, if the mutual information between alleles is positive, indicating selection or non-random mating, the second order entropy representing genotype frequencies will deviate from the value of $2S$ and mean that the heterozygosity will not necessarily be represented by $2pq$.

\subsection{Drift amongst more than two alleles}

As will consistently be shown, one of the powers of the entropy method is that it allows you to generalize results seamlessly under multiple conditions. One case here is the presence of more than two alleles at a locus.  For example, let's look at the 3 allele model where one locus can have alleles $p$, $q$, and $r$ where
$p + q + r  = 1$. Using the log approximation for $S$ we have the following result

\begin{equation}
S \approx p(1-p) + q(1-q) + r(1-r)
\end{equation}

When we see that $1 - p = q + r$ and the corollaries for $q$ and $r$ then

\begin{equation}
S \approx pq + pr + qp + qr + rp + rq = 2pq + 2pr + 2qr
\end{equation}

This is equal to the total ratio of heterozygosity amongst all combinations for the three alleles. Therefore, in the 3 (or $n$) allele model, drift reduces the total ratio of all heterozygous combinations similar to heterozygosity in the two allele combination. This exactly matches the same conclusions reached by Kimura in his analysis of drift in a multi-allelic locus \cite{kimuramulti} where he showed total heterozygosity always decreases at a rate $1/2N_e$ per generation for any number of alleles. The magnitude of each heterozygous combination depends on the allele frequencies of its constituting alleles.

\subsection{The Diffusion Approximation}

Finally, I will show that from equation \ref{driftequation2} you can derive the diffusion approximation first derived by Fisher \cite{fisherdiffuse1,fisherdiffuse2}, expounded on by Wright \cite{wrightdiffuse}, and widely popularized by Kimura \cite{kimurastoch,kimuradiff}. The full derivation of the below will be shown in Appendix A 

\begin{equation}
\frac{\partial f}{\partial t} = \frac{1}{4N_e} \frac{\partial^2}{\partial x^2}f(x)x(1-x)
\label{diffusion10}
\end{equation}

\section{Entropy Increases by Mutation}

The next process we will study is mutation where the mutation rate per site per generation is represented by $\mu$. A general study of the overall nucleotide base entropy of infinite and finite length DNA sequences with single nucleotide polymorphisms (SNPs) was performed by Ma et. al. \cite{mutentropy}. In this paper we will instead look at the mutation rate for alleles and consider the overall entropy introduced to the allele frequency distribution. Mutation introduces genetic diversity and thus is an entropy increasing process. The entropy introduced by mutation is relatively straightforward

\begin{equation}
S_m = -\mu \log \mu  - (1-\mu) \log (1-\mu)
\label{mutation}
\end{equation}

Given the usual low magnitude of $\mu$ on the scale of $10^{-5}$ - $10^{-8}$, this overall effect is small. Taken in isolation,  $\Delta S = S_m$ meaning that every generation there is a constant incremental entropy change linked to a probability $x=\mu$ which corroborates the conclusion that the mutation rate is also the probability of fixation for a mutation in a population.

It is also possible to derive the expected results from the drift-mutation balance and selection-mutation balance. Here we will treat the drift-mutation balance. The overall change in entropy is represented by

\begin{equation}
\Delta S = -\frac{1}{2N_e}S + S_m
\label{driftmutbal1}
\end{equation}

At balance, the entropy of the allele frequency distribution remains constant though the individual alleles are in a dynamic equilibrium. Thus, $\Delta S = 0$ and

\begin{equation}
\frac{1}{2N_e}S = S_m
\label{driftmutbal2}
\end{equation}

Again, we can simplify $S_m$ in a similar manner to that in equation \ref{approximate}, where $S_m \approx 2\mu(1-\mu)$. Given that $ \mu \ll 1$ this can be further simplified to $S_m \approx 2\mu$. Substituting $2\mu$ for $S_m$ in equation \ref{driftmutbal2} and again approximating $S$ as $h$ we derive the steady state heterozygosity at drift-mutation balance for the infinite site model

\begin{equation}
h = 4N_e\mu 
\end{equation}

We will return to a discussion of selection-mutation balance in the next section on mutual information.

\section{Mutual Information: Modeling Selection and Non-Random Mating}

The aforementioned evolutionary effects, despite mutation and drift, assume that random mating and thus the frequency of alleles is the main variable in determining genotype frequencies. Here we will deal with the violation of this assumption, normally caused by selection and inbreeding, which leads to differential survival and reproduction rates amongst genotypes.

Both of these effects, usually given separate treatments, are unified in that they are both causes of increased mutual information between alleles. In other words, the genotype frequencies will not reflect purely random combinations and the allele and genotype frequency will change across generations owing to this.

As shown in equations \ref{MI} and \ref{MI2}, the genotype takes center stage in mutual information. Here we demonstrate the effect of mutual information on both allele frequencies and genotypes. The important quantity is the mutual information between alleles $p$ and $q$ between two generations due to effects of selection or non-random mating. Following this,

\begin{equation}
I(t,t-1) = S_t + S_{t-1} - S_t(p,q)
\label{basicMIequation}
\end{equation}

to find the allele entropy change, we manipulate this equation with $2S_{t-1}$

\begin{equation}
I(t,t-1) - 2S_{t-1} = S_t - S_{t-1} - S_t(p,q)
\label{basicMIequation2}
\end{equation}

and

\begin{equation}
\Delta S = S_t - S_{t-1} = I(t,t-1) + S_t(p,q)- 2S_{t-1}  
\label{basicMIequation3}
\end{equation}

Unlike the Kullback-Leibler divergence, mutual information increases the entropy, however, combined with the $-2S_{t-1}$ the overall effect of selection is usually negative reducing the overall diversity in the population except in cases where a relatively rare allele has a selective advantage. The change in entropy is also related to the value of the joint entropy between $p$ and $q$. When $S_t(p,q)$ is represented by HW proportions and no other evolutionary forces are acting, the right hand side reduces to 0 and the population has a constant allele frequency. If the value $2S_{t-1}-S_t(p,q)$ is considered as a type of quasi-mutual information, $I'$, between the allele frequencies in $t-1$ and the genotype frequencies in $t$ then equation \ref{basicMIequation3} can be restated as

\begin{equation}
\Delta S = S_t - S_{t-1} = I - I'
\label{basicMIequation4}
\end{equation}

This combination is not easily analytically tractable under most circumstances. $I'$ is not a formally defined quantity and differs from mutual information in many aspects, one of which is that it can be negative. It is mainly used for conceptual and notational convenience. However, in most cases, $S_{t-1} > S_t$ since there is an overall decrease in entropy between generations as the overall distribution of allele frequencies is changed by selection. Therefore, often times $I' \gg I$ so that

\begin{equation}
\Delta S \approx -I'
\label{basicMIequation5}
\end{equation}

$I'$ can typically be reduced by to the following form where $p' = p^2\frac{w_{11}}{\bar{w}} + pq\frac{w_{12}}{\bar{w}}$ and $q' = q^2\frac{w_{22}}{\bar{w}} + pq\frac{w_{12}}{\bar{w}}$

\begin{eqnarray}
I' &=& -2(p-p')\log p - 2(q-q')\log q \nonumber \\
&+& p^2\frac{w_{11}}{\bar{w}}\log \frac{w_{11}}{\bar{w}} \nonumber \\
&+& 2pq\frac{w_{12}}{\bar{w}}\log \frac{w_{12}}{\bar{w}} \nonumber \\
&+& q^2\frac{w_{22}}{\bar{w}}\log \frac{w_{22}}{\bar{w}}\nonumber \\
\end{eqnarray}

or

\begin{eqnarray}
I' &=& -2(p-p')\log p - 2(q-q')\log q \nonumber \\ 
&+& p^2\frac{w_{11}}{\bar{w}}\log \frac{w_{11}}{\bar{w}} \nonumber\\ 
&+& 2pq(1-hs)\frac{w_{11}}{\bar{w}}\log (1-hs)\frac{w_{11}}{\bar{w}}\nonumber\\ 
&+& q^2(1-s)\frac{w_{11}}{\bar{w}}\log (1-s)\frac{w_{11}}{\bar{w}} \nonumber \\
\label{selecI}
\end{eqnarray}

We can now solve for two cases of mutation-selection balance where $0 < s \leq 1$ and $h < 1/2$ \cite{burger} and the equilibrium is asymptotically stable. Equation \ref{selecI} can be used to interpret the balance conditions for selection and mutation by setting $I' = 2\mu$. For example, take the case where the allele $q$ is selected against with a strength measured by $s$ and $p \approx 1$ as well as $w_{11} \approx \bar{w} \approx 1$. In addition, given $q$ is the rare allele, $q-q' \approx \mu$. First where $h=0$

\begin{equation}
- 2\mu \log q + q^2(1-s)\log (1-s)=2\mu
\end{equation}

and

\begin{equation}
q^2(1-s)\log (1-s) = 2\mu(1 + \log q)
\end{equation}

using the log approximation for $1-s$ and approximating $1 \gg s$

\begin{equation}
q^2 = -\frac{2\mu(1 + \log q)}{s}
\end{equation}

given that $q$ is likely very small, $\log q$ is a correspondingly large negative value. If $|\log q| > 1$ but the differences in orders of magnitude between $\mu$ and $s$ are large we can come up with the familiar approximation of
\begin{equation}
q \approx \sqrt{\frac{\mu}{s}}
\end{equation}

Similar procedures apply when $h>0$ and the heterozygous frequency is the dominant presence of the recessive allele

\begin{equation}
2q(1-hs)\log(1-hs) = 2\mu(1 + \log q)
\end{equation}

and assuming $1\gg hs$ and with similar arguments regarding $\log q$

\begin{equation}
q \approx \frac{\mu}{hs}
\end{equation}

In conclusion, just like drift, the main conclusions of population genetics can be readily derived using the information theoretic description. 

\section{Entropy Changes due to Gene Flow}

The final change in allele frequency we will deal with is the change in entropy by the migration of a population to or from the one under analysis. For example here, we will base our results on the simple Wright mainland-island model. To determine the total entropy change in the recipient (island) population, we must calculate the combined entropy of both the resident identical by descent population and the immigrants. For the island population with allele frequencies $p$ and $q$ which has a percentage of its population, $m$, as migrants with allele frequencies for the same alleles of $p^*$ and $q^*$ we can calculate the combined entropy as below,

\begin{equation}
S_{total} = S_{migrants} + S_{island}
\end{equation}

\begin{eqnarray}
S_{total} & = & \nonumber \\
&-&\frac{p^*m2N_e}{2N_e}\log \frac{p^*m2N_e}{2N_e} \nonumber \\  
&-&\frac{q^*m2N_e}{2N_e}\log \frac{q^*m2N_e}{2N_e}\nonumber \\
&-&\frac{p(1-m)2N_e}{2N_e}\log \frac{p(1-m)2N_e}{2N_e}\nonumber \\
&-&\frac{q(1-m)2N_e}{2N_e}\log \frac{q(1-m)2N_e}{2N_e}\nonumber\\
\end{eqnarray}

\begin{equation}
-p^*m\log p^*m - q^*m\log q^*m - p(1-m)\log p(1-m) - q(1-m)\log q(1-m)\\
\end{equation}

The logarithms can then be expanded to produce

\begin{eqnarray}
&-&p^*m\log p^* - p^*m\log m -q^*m\log q^* - q^*m\log m\nonumber \\
&-&p(1-m)\log p - p(1-m)\log (1-m) - q(1-m)\log q - q(1-m)\log (1-m)\nonumber\\
\end{eqnarray}

\begin{eqnarray}
&-&m(p^*\log p^*+ q^*\log q^*) - m\log m(p^* + q^*) \nonumber \\
&-&(1-m)(p\log p + q\log q) - (1-m)\log (1-m) (p+q)\nonumber\\
\label{geneflow4}
\end{eqnarray}

Given that both $p^* + q^*=1$ and $p + q = 1$ and defining $S = -p\log p - q\log q$ and $S^* = -p^*\log p^* - q^*\log q^*$ we  finally reduce equation \ref{geneflow4} to

\begin{equation}
S_{tot} = S + m(S^*-S) - m\log m - (1-m)\log (1-m) 
\end{equation}

Here we see a fortuitous derivation. With $m$ being the ratio of the population from migrants, we see that we have derived the basic formulation for Gibbs' entropy of mixing in 

\begin{equation}
S_{mix} = - m\log m - (1-m)\log (1-m)
\end{equation}

so our final expression for the total entropy change due to gene flow is

\begin{equation}
S_{tot} = S + m(S^*-S) + S_{mix} 
\end{equation}

\begin{equation}
\Delta S = m(S^*-S) + S_{mix}  
\end{equation}

Note the entropy change varies with $m(S^*-S)$ which is an exact analogue of the change in probability $p$ during gene flow where $\Delta p = m(p^* - p)$. In the case where the immigrating populations have the same allele frequency distribution as the island population $S^*=S$ and we reduce to

\begin{equation}
\Delta S = S_{mix}
\end{equation}

This raises a paradox similar to the one Gibbs confronted about 140 years ago in the theory of statistical mechanics. In this case, the amalgamation of two distinct populations can be modeled in a similar manner to the entropy change of mixing in thermodynamic processes where the entropy of mixing and the weighted average entropy of the two populations combine to determine the entropy of the new combined population.

However, this raises a paradox. On one hand we have the classical result  $\Delta p = m(p^* - p)$ so for populations with identical allele frequencies $\Delta p = 0$. On the other hand, the increase in entropy driven by the entropy of mixing directly predicts a change in the overall entropy which necessitates a change in the allele frequencies. The solution, when the proportion of immigration equals that of death and emigration, is that the entropy of mixing is offset by the entropy of ``de-mixing'' when the proportion $m$ of the population from the previous generation dies or migrates away as the model implies. If you replace $p^*$ and $q^*$ in equation \ref{geneflow4} with  $p$ and $q$ we see that

\begin{equation}
S_{total} = S + S_{mix}
\end{equation}

so the proportion of the population leaving would cause a ``de-mixing'' of exactly the same magnitude. Therefore, when two populations with identical allele frequencies mix and the population size stays constant, $\Delta S = 0$ as expected. For two populations with differing allele frequencies

\begin{equation}
\Delta S = m(S^*-S)
\end{equation}

Next, we will look at drift-migration balance.

\begin{equation}
0 = -\frac{1}{2N_e}S + m(S^* - S)
\end{equation}

and

\begin{eqnarray}
S = \frac{2Nm}{1 + 2Nm}S^* \nonumber\\
h = \frac{2Nm}{1 + 2Nm}h^* \nonumber\\
\end{eqnarray}

which implies a fixation index, $F_{st}$ of

\begin{equation}
F_{st} = \frac{1}{1+ 2Nm}
\end{equation}

This shows that under the simple model we find that the observed heterozygosity is equal to the heterozygosity of the migrating population times $\frac{2Nm}{1 + 2Nm}$ which can be an approximation of $1 - F_{st}$. If $h^*$ is the original heterozygosity of the island population $h$, we can see the standard balance for drift and migration for small $m$ becomes the observed heterozygosity equaling the expected heterozygosity times $1 - F_{st}$.







\section{Master Balance Equation}

From the foregoing discussions and given that entropy is an extensive quantity whose total amount is additive, we can begin to look at the entire evolution of a population's allele and genotype frequencies in better detail. Specifically the full equation for changes in the allele frequency is given by

\begin{equation}
\Delta S = -D +(I-I') + S_m + S_f
\label{mastereq1}
\end{equation}

where $S_f$ represents the change in entropy due to gene flow. As will be shown in the simulation results in the next section, the change in the allele frequencies in a population due to all evolutionary forces can be simulated using entropy and matched with simulated results. In addition, we can obtain a master balance equation subject to all forces when $\Delta S = 0$ by showing the following

\begin{equation}
D + I'-I = S_m + S_f
\label{mastereqequil1}
\end{equation}

expanding assuming that $I'\gg I $ and $S_m=2\mu$

\begin{equation}
\frac{1}{2N_e}S_{t-1} + 2S_{t-1} - S_t(p,q) = 2\mu + m(S^* - S_{t-1})
\label{mastereqequil2}
\end{equation}

\begin{equation}
S_{t-1}\bigg(2+\frac{1}{2N_e} + m\bigg) = 2\mu + S_t(p,q) + mS^*
\label{mastereqequil3}
\end{equation}

If in the case of no or balanced gene flow ($S^*=S$) we can reduce to

\begin{equation}
S_{t-1}\bigg(2+\frac{1}{2N_e}\bigg) = 2\mu + S_t(p,q)
\label{mastereqequil3.5}
\end{equation}

and finally substituting $h$ for $S_{t-1}$ in equation \ref{mastereqequil3} when there is no mutual information

\begin{equation}
h\bigg(\frac{1}{2N_e} + m\bigg) = 2\mu + mS^*
\label{mastereqequil4}
\end{equation}

The terms on the left side largely represent population size (extensive) effects while those on the right side represent intensive effects on allele frequencies due to mutation and selection. This gives a general equation for measuring either the change in heterozygosity or the change in other evolutionary parameters over a timescale where allele frequencies are relatively stable. 

\subsection{Changes in Entropy over Multiple Generations}

Obviously, there often may be a situation calling for the analysis of the change in entropy across multiple generations. Given the preceding equations, this can be done iteratively using computer simulation (as will be demonstrated in the next section) or in some limited cases, analytically. In particular, one can look at the entropy several generations into the future or past if certain assumptions are made regarding the stationarity of certain parameters.

The easiest assumptions to deal with are genetic drift and mutation. By analyzing the master equation iteratively only involving drift and mutation, one can calculate the entropy $S_t$ given the entropy $d$ generations in the past $S_{t-d}$ with the following equation

\begin{equation}
S_t = S_{t-d}\bigg(1-\frac{1}{2N_e}\bigg)^d + 2\mu\sum_{i=0}^{d-1} \bigg(1-\frac{1}{2N_e}\bigg)^i
\end{equation}

For a large number of generations the first term goes towards zero and the second term geometric series converges to $2N_e$ giving

\begin{equation}
S_t = 4N_e\mu
\end{equation}

Selection can be similarly integrated into the analysis, however, given the sometimes volatile nature of selection and how it integrates competition, mutualism, environment, and disease among other variables, it is questionable whether a steady state relative selection coefficient bears much semblance to reality.

\subsection{Boundary Conditions}

A final key feature that we must understand is the behavior of the entropy evolution equations at the two boundaries of minimum and maximum entropy, 0 and $\log n$ respectively. At $S=0$, one allele becomes fixed and the other is lost and therefore both genetic drift and mutual information disappear. Given that additional mutation and gene flow only act to increase entropy, then the behavior of the equations is completely consistent at the boundary at $S=0$. For maximum entropy, by definition $\Delta S = 0$ and therefore, even though there will likely be mutation and migration which push to increase entropy, there must be counterbalancing effects. Maximum entropy conditions can only arise in the absence of any selective pressure or nonrandom mating on allele frequencies or genotypes and mutation and gene flow must balance with genetic drift. However, one could easily imagine a hypothetical situation of a large population where genetic drift is negligible over appreciable time scales but entropy increases due to mutation and migration push the entropy to its maximum. In the next generation, it would seem the master evolution equation would dictate that the entropy must increase above $\log n$ in the subsequent generation. However, one must understand when allele frequencies are perfectly balanced at maximum entropy, incremental mutation or mixing from gene flow must necessarily reduce the entropy below its maximum value and therefore despite the general nature of the equations, $S \leq \log n$. Therefore, at the boundaries we should define 
\begin{eqnarray}
S = \log n & S_m=-2\mu \nonumber \\
\end{eqnarray}

\section{Simulation Results}

Throughout much of the paper, it has been asserted that the methods based on information theoretic quantities were as effective as those from theoretical population genetics models. This section will test that assertion by running a 1000 trial Monte Carlo of the evolution of the allele frequencies and entropy of a 250 member population over 1000 generations. The results of the simulation will then be compared with the predicted evolution of the population using the techniques derived in this paper and using the same fixed parameters. Note that all of these simulations only use the information theoretical parameters and not the Mercator approximation.

In the following figures are comparisons for both the frequency of allele $p$ and the entropy $S$ where the solid green line is the simulation output, using the Python version of SimuPOP, and the dashed red line is the output from the information theoretic method. For the example of pure drift-mutation, the heterozygosity proportion will replace $p$ on the y-axis.

\begin{table*}
\centering
\begin{tabular}{cc}
\includegraphics[width=2.0in,height=2.0in]{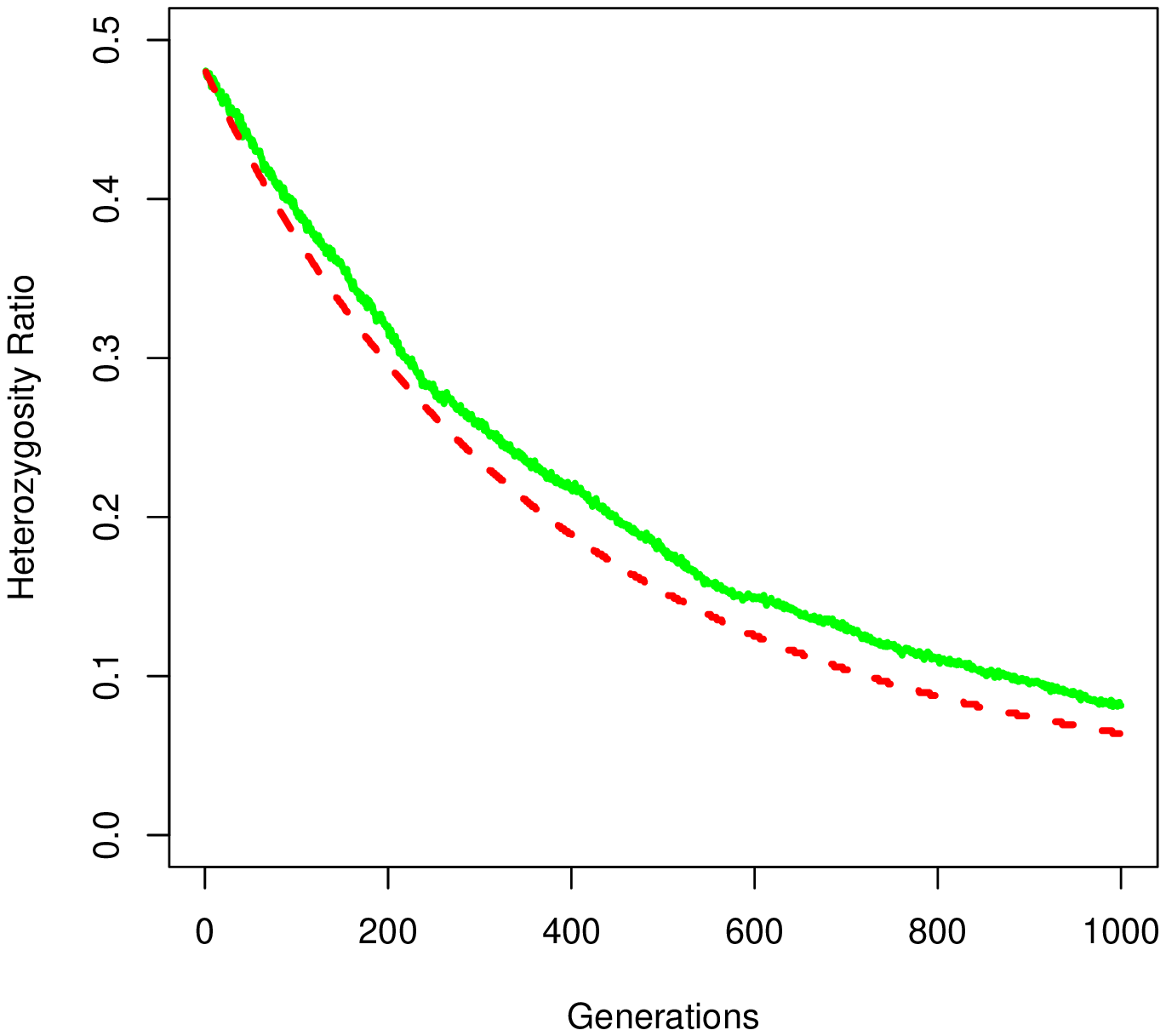}&\includegraphics[width=2.0in,height=2.0in]{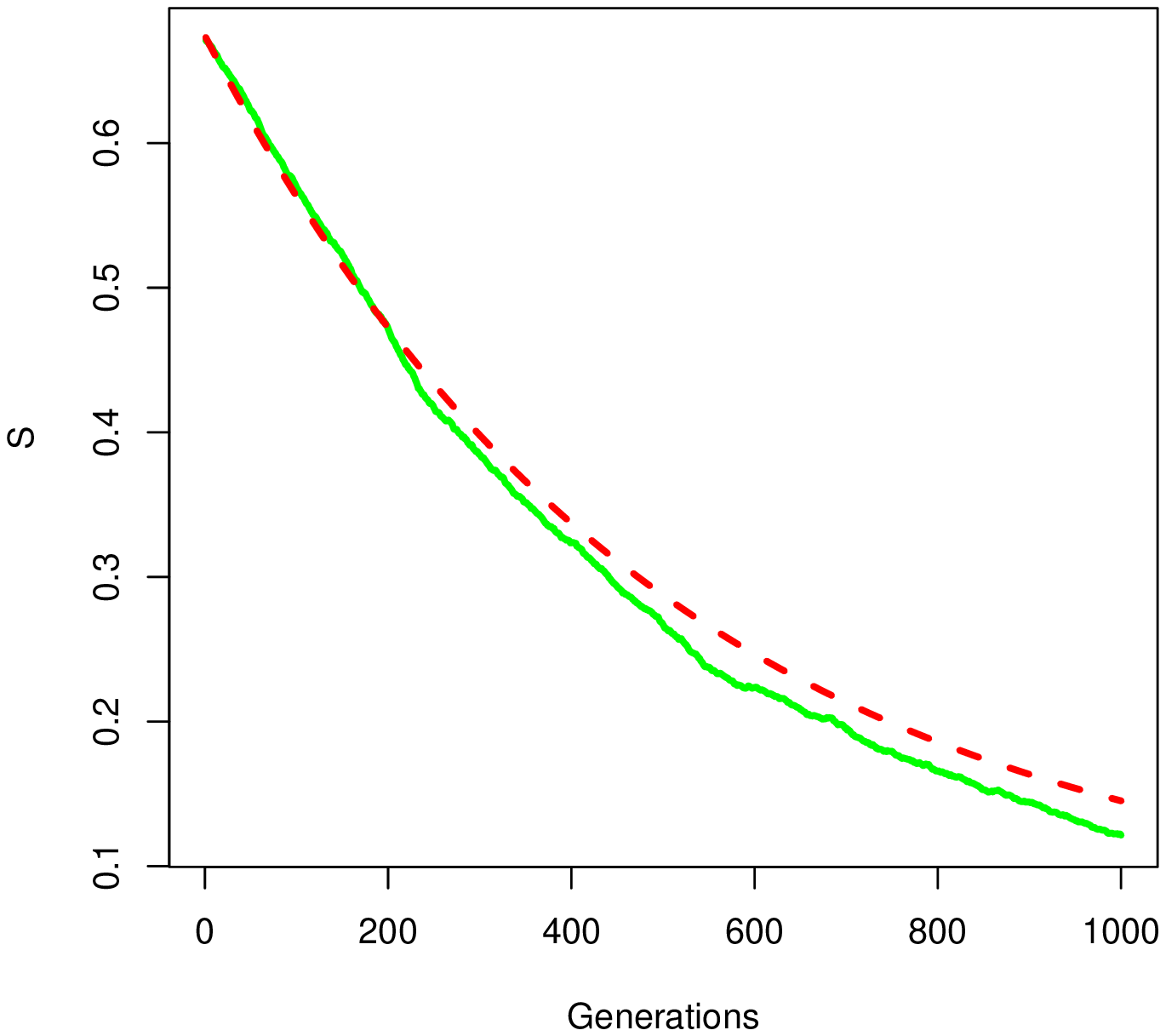}\\
\includegraphics[width=2.0in,height=2.0in]{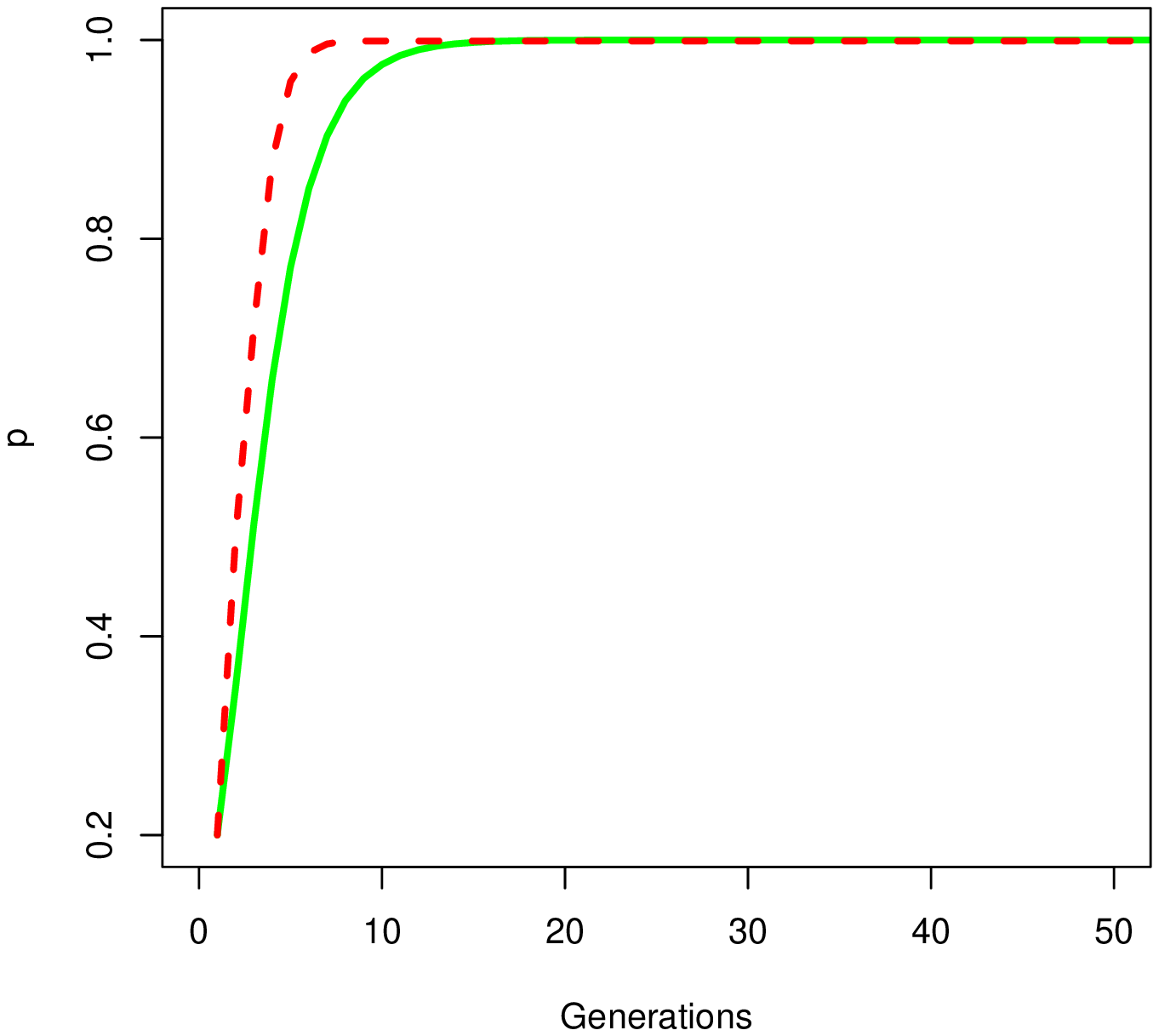}&\includegraphics[width=2.0in,height=2.0in]{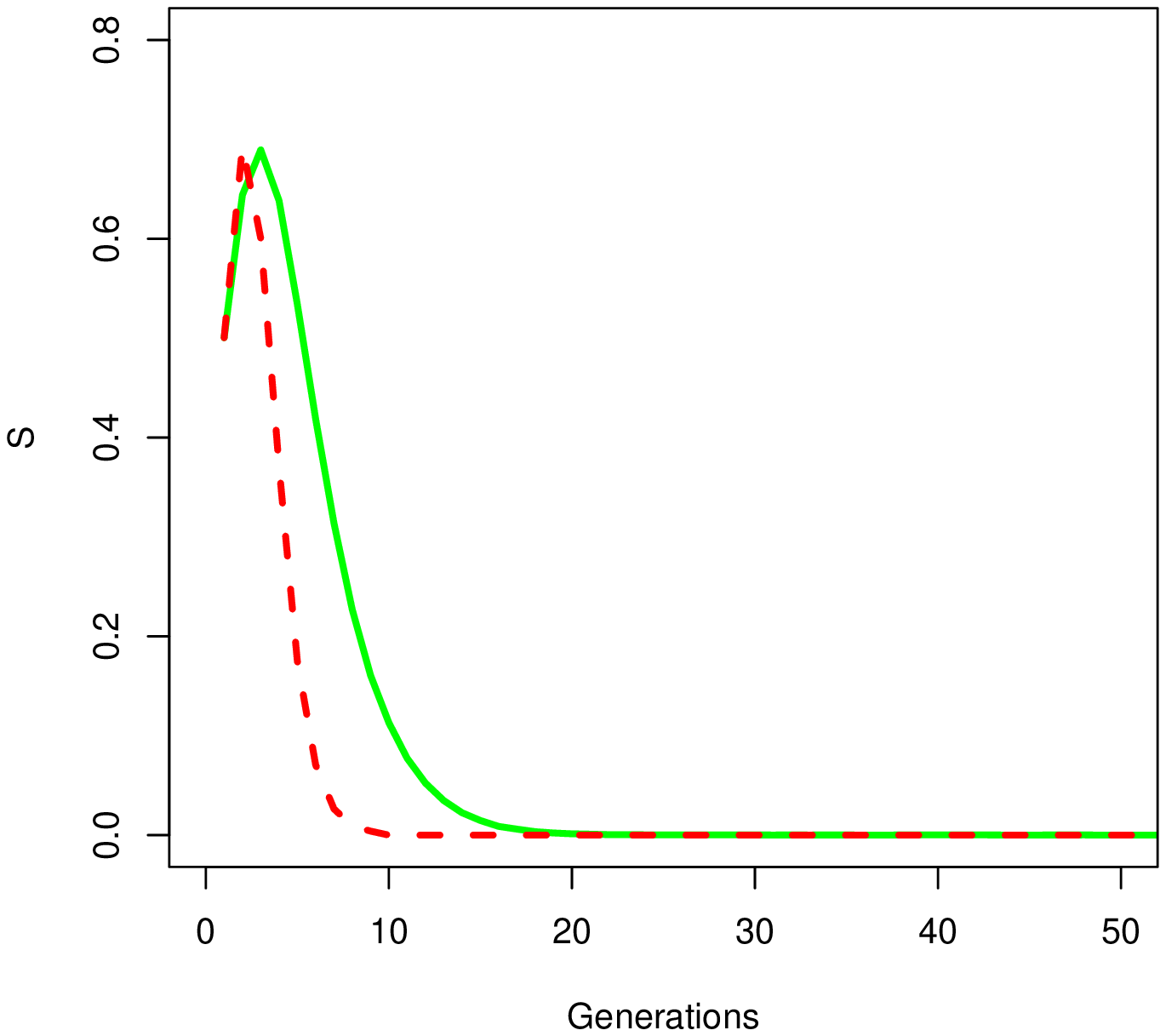}\\
\includegraphics[width=2.0in,height=2.0in]{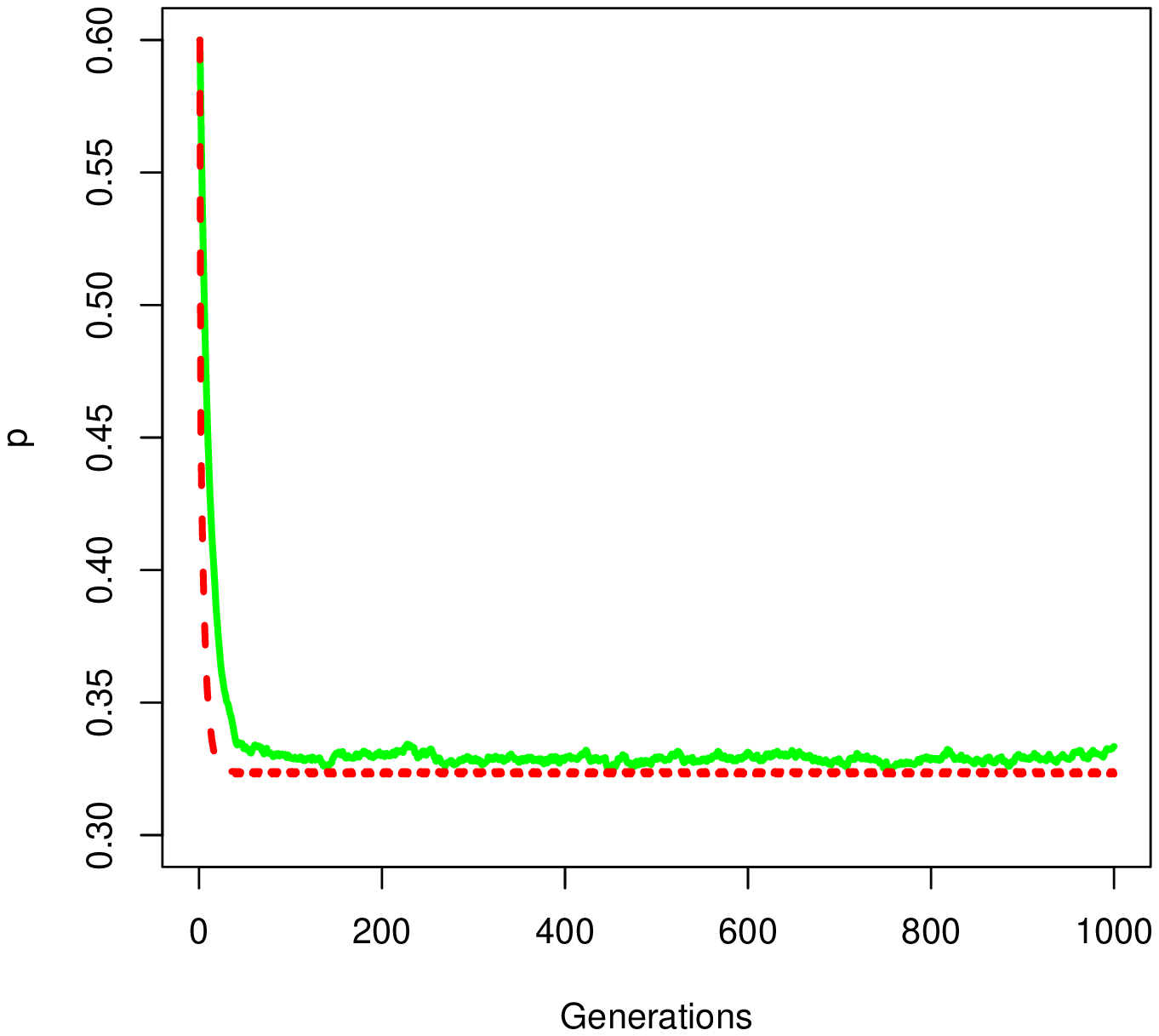}&\includegraphics[width=2.0in,height=2.0in]{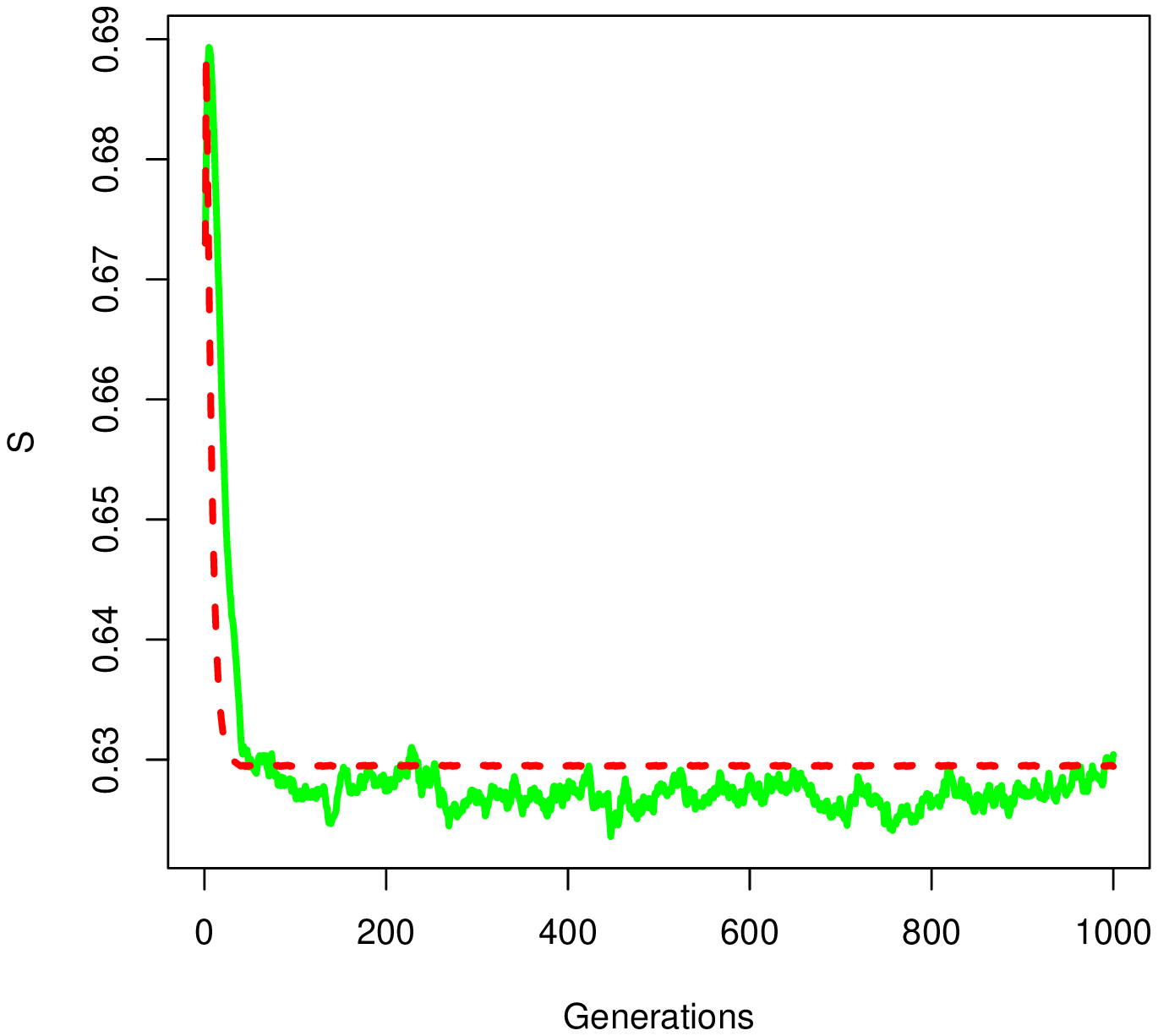}\\
\end{tabular}
\caption{Simulation and entropy method results for three cases of population evolution with fixed parameters. Each simulation shows a population of $N_e = 250$ with a mutation rate $\mu = 10^{-5}$ over 1000 generations. Each figure on the left side is the frequency of allele $p$ over time with the exception of the first figure which is the heterozygosity ratio. Each figure on the right is the allele frequency entropy over time. The first pair represents drift-mutation with no other evolutionary forces acting with starting frequencies $p =0.6, q=0.4$. The second pair represents classic incomplete dominance with initial allele frequencies $p =0.2, q=0.8$ and $s=0.75,h=0.49$. The third pair is overdominance with starting values of $p=0.6,q=0.4$ and $s=-0.14,h=2$.}
\label{simtable}
\end{table*}
\newpage
\begin{table*}
\centering
\begin{tabular}{cc}
\includegraphics[width=2.0in,height=2.0in]{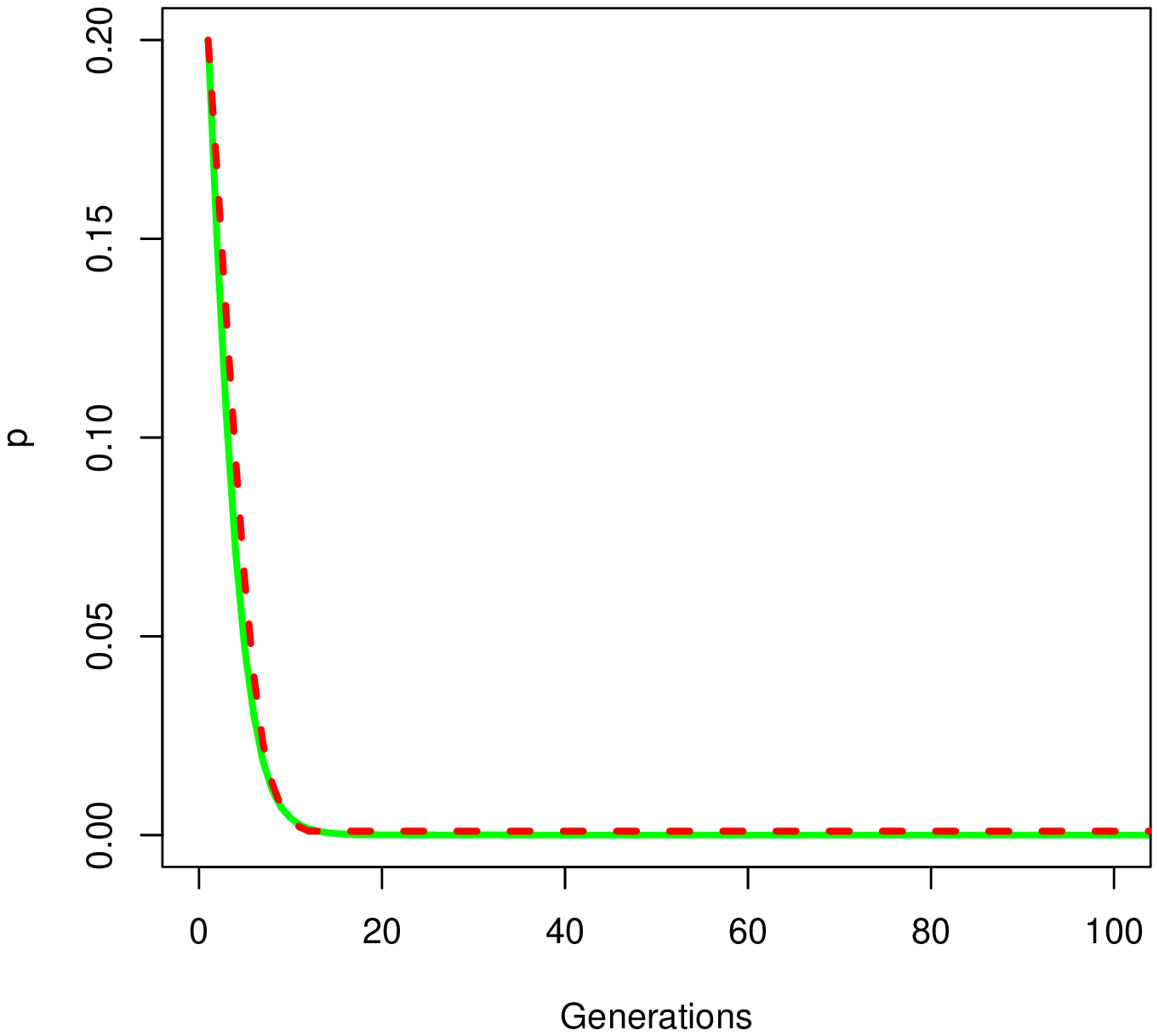}&\includegraphics[width=2.0in,height=2.0in]{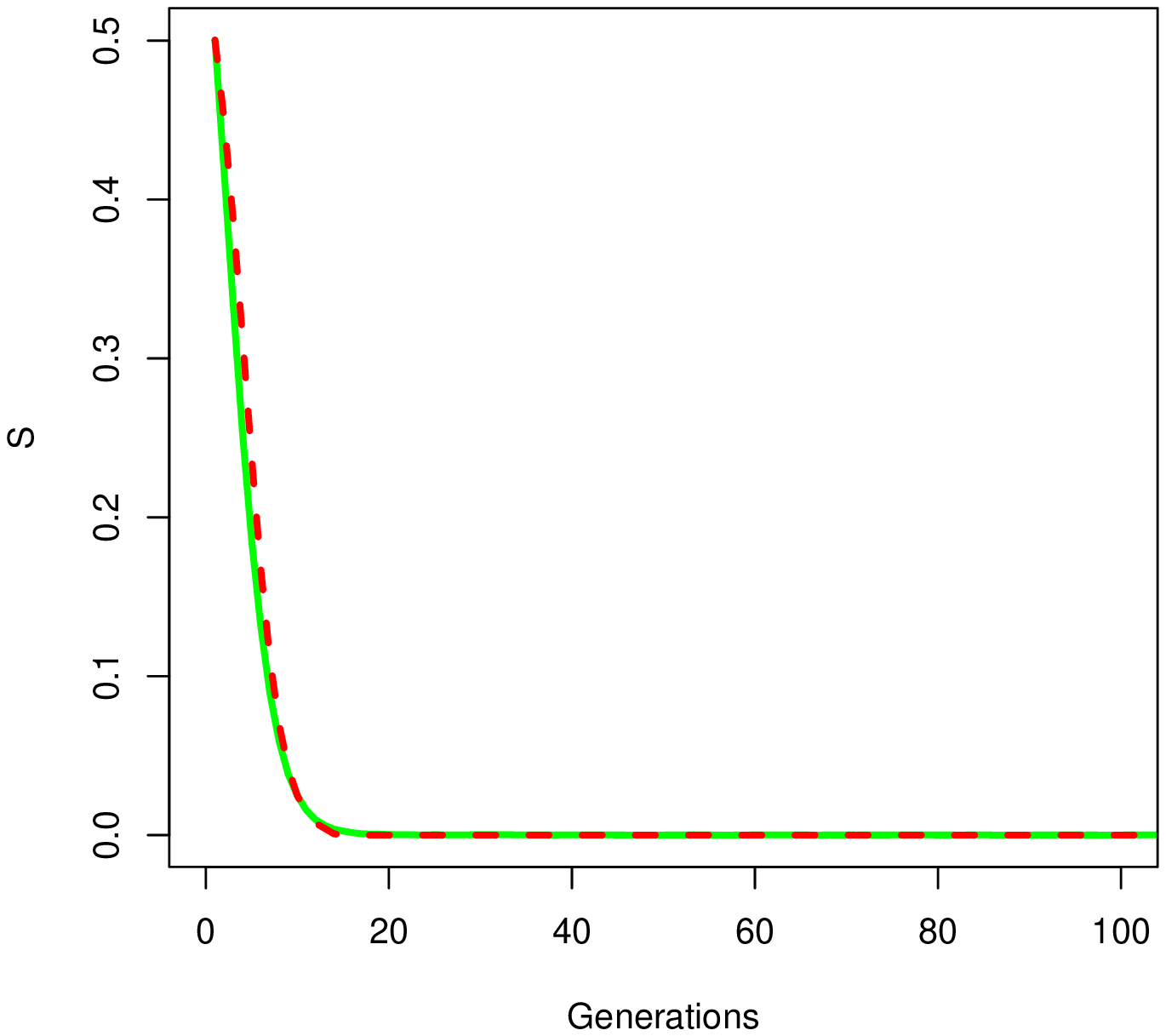}\\
\includegraphics[width=2.0in,height=2.0in]{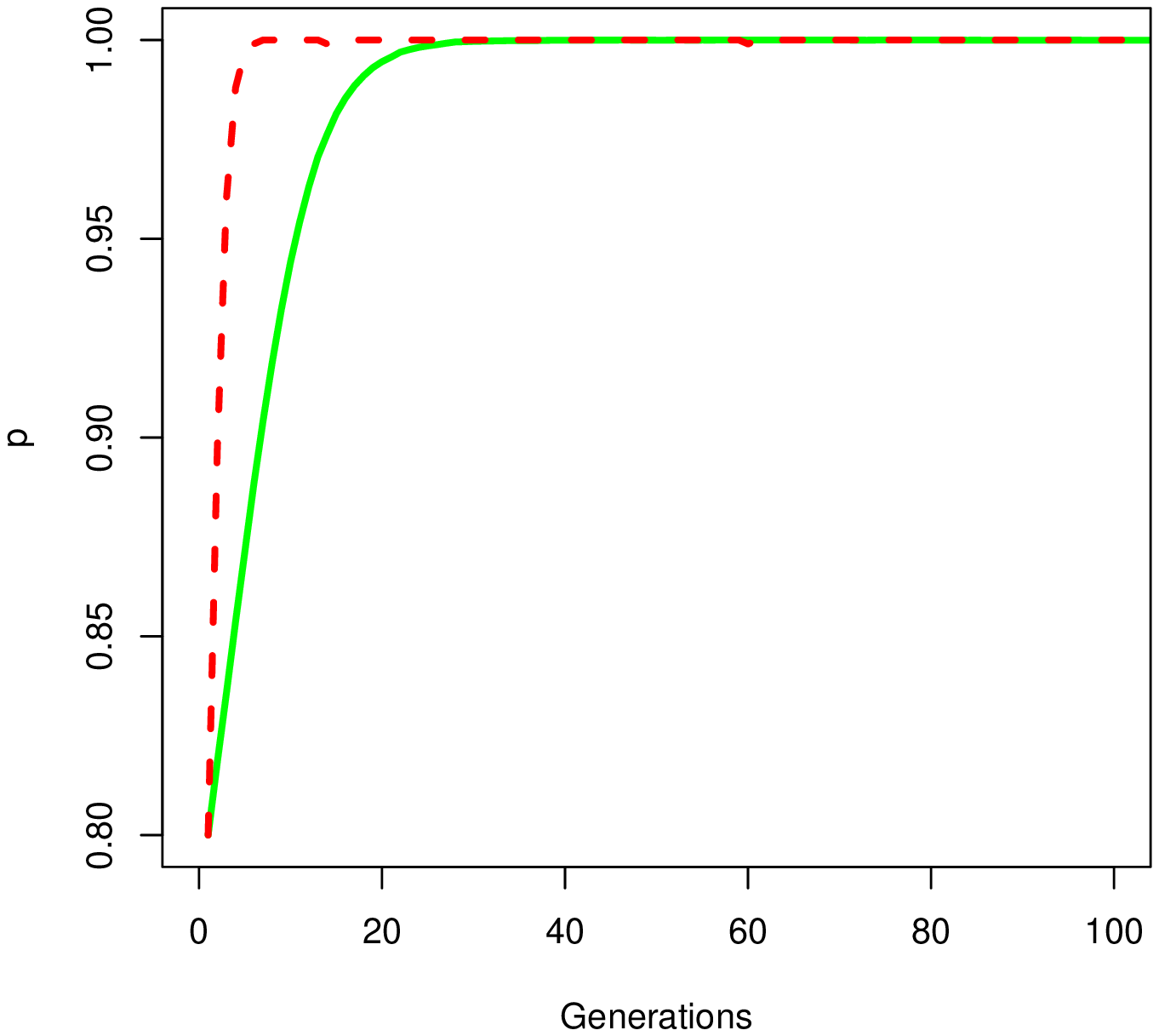}&\includegraphics[width=2.0in,height=2.0in]{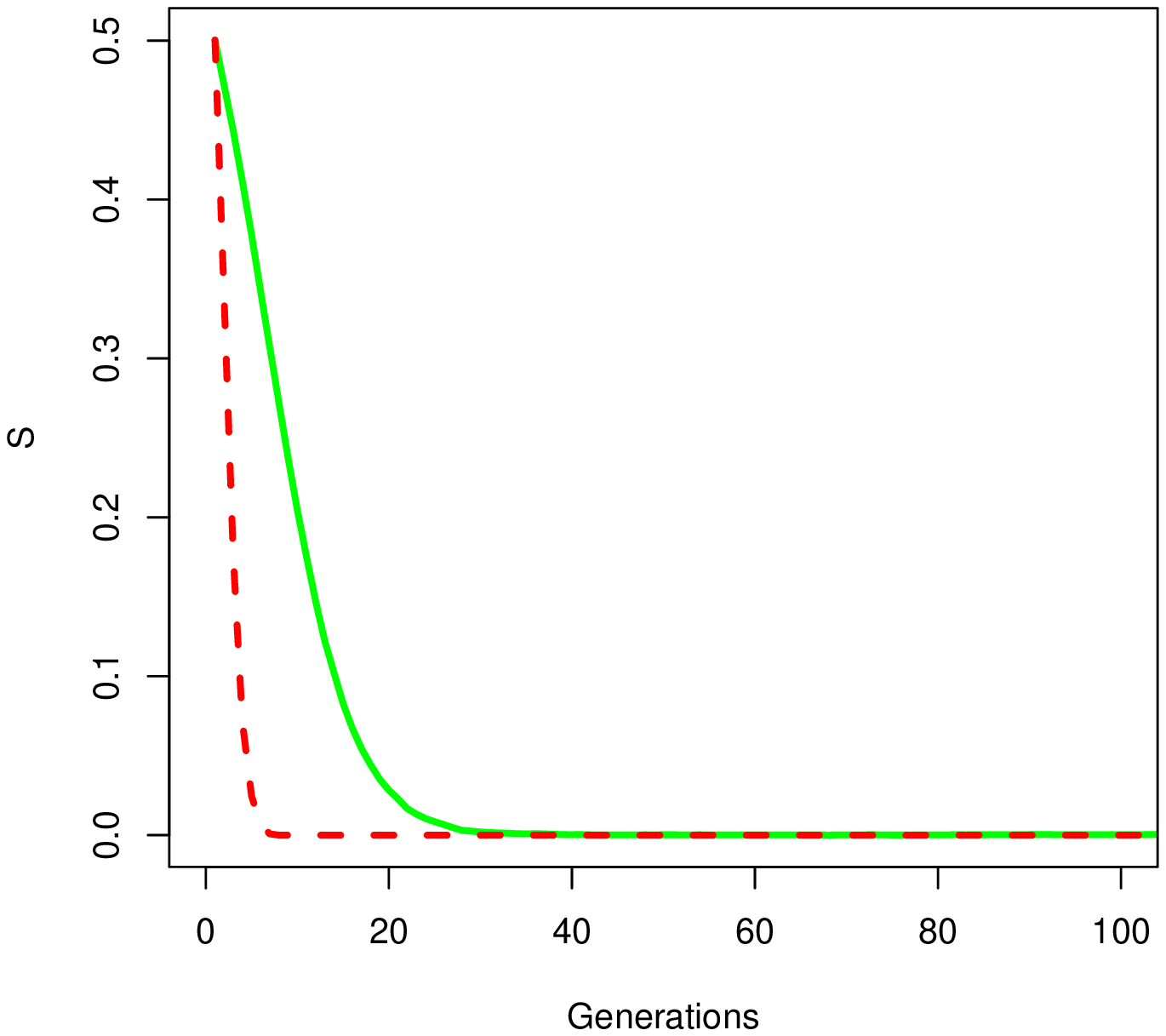}\\
\end{tabular}
\caption{Simulation and entropy method results for two cases of underdominance with different starting frequencies for $p$. Each simulation shows a population of $N_e = 250$ with a mutation rate $\mu = 10^{-5}$ over 1000 generations. Each figure on the left side is the frequency of allele $p$ over time. Each figure on the right is the allele frequency entropy over time. Both pairs represent underdominance with fitness variables $s=0.2,h=2$. The first pair have starting frequencies $p=0.2,q=0.8$ and the second pair have starting frequencies  $p=0.8,q=0.2$. Note that in both cases, though the evolution of $p$ is sensitive to its initial value, the value of $S$ is approximately the same.}
\label{simtable2}
\end{table*}


\section{Multiple Loci Models and Linkage Disequilibrium}

One of the key flexibilities of the information theoretical method is that it can be easily expanded to investigate systems with multiple alleles and multiple loci, even if the probabilities or outcomes are not analytically tractable. As you expand the analysis amongst multiple loci mutual information calculations become more important and take center stage versus almost all other considerations. A key example is the model of linkage disequilibrium. The standard measure of disequilibrium, $D$ \cite{linkdis,linkdis2}, between two loci with alleles, $A$ and $B$ is

\begin{equation}
D = P(A,B)-P(A)P(B)
\label{LinkD}
\end{equation}

Other measures of linkage disequilibrium have already been devised using entropy \cite{entropylink}, Kullback-Leibler divergence \cite{entropylink2} and mutual information \cite{entropylink3}.

Unlike the coefficients of relative fitness, however, $D$ is the measure of underlying deviation from equilibrium, not a coefficient for a causal agent of that disequilibrium. Therefore, we can derive $D$ from the mutual information from each other amongst multiple loci. For the case of two loci, it is easy to derive $I$ from $D$. First, for a locus with alleles $A,a$ and $B,b$, 

\begin{eqnarray}
& &D = P(A,B) - P(A)P(B)\\
& &D + P(A)P(B) = P(A,B)\\
& &D_{A,B}=D_{a,b}=-D_{A,b}=-D_{a,B}\nonumber \\
\end{eqnarray}

From this we can derive an expression for the mutual information 

\begin{eqnarray}
I &=& (D + P(A)P(B))\log \bigg(\frac{D}{P(A)P(B)} + 1\bigg)\nonumber\\ 
&+& (-D + P(A)P(b))\log \bigg(\frac{-D}{P(A)P(b)} + 1\bigg)\nonumber\\
&+& (-D + P(a)P(B))\log \bigg(\frac{-D}{P(a)P(B)} + 1\bigg)\nonumber\\ 
&+& (D + P(a)P(b))\log \bigg(\frac{D}{P(a)P(b)} + 1\bigg)\nonumber \\
\end{eqnarray}

Using the log approximation, and multiplying out we can reduce the above to

\begin{equation}
I \approx D^2\bigg[\frac{1}{P(A)P(B)}+\frac{1}{P(A)P(b)}
+\frac{1}{P(a)P(B)}+\frac{1}{P(a)P(b)}\bigg]
\end{equation}

And finally

\begin{equation}
I \approx \frac{D^2}{P(A)P(a)P(B)P(b)}
\label{Ir2}
\end{equation}

Equation \ref{Ir2} is also the exact equation for the alternate measure of linkage disequilibrium known as $r^2$. This shows that under linear approximation $I \approx r^2$ and therefore is a roughly equivalent measure at the two loci level. This fact has previously been discussed in papers on LD and entropy \cite{entropylink} and mutual information \cite{entropylink2}. Mutual information does have an advantage, however, as you increase the number of loci, in that first, the mutual information is not a measure of linear dependence like the correlation coefficient. Second, it can consolidate into one metric the strength of the total relationship amongst all loci. Third and finally, it can be used to measure and compare the relative disequilibrium at different numbers of loci using the multivariate mutual information or interaction information (which can be negative). More work needs to be performed, however, to make sure it is a robust and clear measure of LD with its own advantages versus other measures \cite{gameticcritique1, gameticcritique2}.

A full exposition of the entropy method applied to multiple loci is beyond the scope of this paper, however, we can quickly show that the increased gene diversity that genetic hitchhiking is often used to explain can also be explained using this method.

For a model where we have two loci of allele pairs, $A,a$ and $B,b$, the expected second order entropy of the system can be represented as

\begin{equation}
S(A,B) = S_A + S_B - r^2
\end{equation}

The change in entropy due to selection can be approximated with $I$ and $I'$ similarly where

\begin{equation}
\Delta S = I - I' = r^2 - I'
\end{equation}

Therefore, since $r^2$ is always positive, any level of linkage disequilibrium offsets the rate of reduction of genetic diversity across sites by slowing the change in the entropy decrease caused by selection.

\section{How Useful is Entropy?}

\begin{figure}[ht]
  \centering
  \includegraphics[width=2in,height=2in]{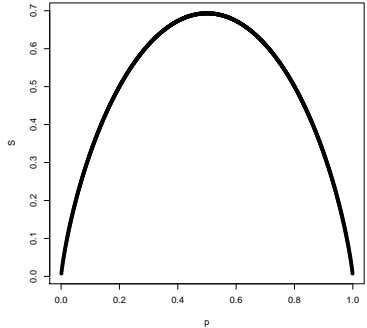}
  \caption{Plot of first order entropy $S$ vs. $p$.}
	\label{entropyplot}
\end{figure}

One aspect of the paper left unmentioned is how we can go from values of entropy to the allele frequencies. For a two allele model, this is relatively simple given that you can do a seek on values of $p$ whose entropy will match the calculated entropy given allowable tolerance. However, one will note that for an entropy function (see figure \ref{entropyplot}) there are two possible $p$ values for every value of entropy, one for $p$ and one for $1-p$. The solution to this is to have the software track the starting allele frequency for $p$ and using the subsequent entropy change, you can determine the direction in the change of $p$ by whether $p\geq 0.5$ or $p<0.5$. Since the difference equation is equivalent to a first order differential equation, the entropy function has a monotonic increase or decrease in any given time step and cannot ``skip'' one solution past the maximum entropy to another of the same entropy in the same time step. Therefore, the closest value of $p$ to the starting value which fulfills the criterion of the change in entropy is the solution.

This highlights one of the key weaknesses of the method based of entropy and information theory: it can only calculate the structure of the distribution and does not differentiate between which alleles take what values in the frequency distribution. Therefore, any $n$ allele model can give $n!$ different possible matches between alleles and frequencies. This can only be distinguished by selecting one of the results as the most biologically feasible, often given the assumptions of the parameters for selection or non-random mating in the mutual information. Also, the overall entropy does not distinguish between alleles which are identical by descent, just the aggregate distribution.

Finally, though the author again cautions against undue wild speculation on the connections between evolutionary processes and information theory, it is not inappropriate to note that such links may help us to understand more profound connections between biological processes and information theory.

On the speculative side, a possible interesting result of this formalism would be an enhanced understanding of other types of evolution in systems that are not biological, but exhibit similar characteristics of discrete hereditary units which undergo forces that can be represented by the same information theoretic parameters. Be they artificial life simulations, malicious code evolution, or some other unimagined paradigm, they would be able to display similar evolution to that we observe in nature without having the same basic constitution, underlying biology, genetic coding or inheritance mechanisms, or even organic compounds. The universality of evolutionary processes could be deeper than we realize. A short excursion in this light is given in Appendix B.

In conclusion, this paper has endeavored to show that the biological forces of evolution can be linked to an information theoretic representation that devises a comprehensive equation based on entropy that reproduces the commonly known features of evolutionary change in allele frequencies and genotypes. Whether this method will only reproduce what is already known in population genetics or produce new and unexpected insights that can be validated through genomic data will be an interesting question to be answered in future works.

\newpage
\appendix
\section{The Diffusion Approximation}

The information theory techniques used to model genetic drift can be used to approximate diffusion. 

Starting from pure drift we have

\begin{equation}
\Delta S = -\frac{1}{2N_e}S
\end{equation}

with its continuous time version of

\begin{equation}
\frac{\partial S}{\partial t} = -\frac{1}{2N_e}S
\label{diffusion1}
\end{equation}

From the definition of entropy being $S = - \int^\infty_{-\infty} f(x) \log f(x) dx$ we can transform equation \ref{diffusion1} to

\begin{equation}
- \int^\infty_{-\infty} \frac{\partial}{\partial t}f(x) \log f(x) dx = \frac{1}{2N_e} \int^\infty_{-\infty} f(x) \log f(x) dx
\label{diffusion2}
\end{equation}

\begin{equation}
\frac{\partial}{\partial t}(f(x) \log f(x)) = -\frac{1}{2N_e} f(x) \log f(x)
\label{diffusion3}
\end{equation}

\begin{equation}
(1+\log f(x))\frac{\partial f}{\partial t} = -\frac{1}{2N_e} f(x) \log f(x)
\label{diffusion4}
\end{equation}

Next we can simplify this equation by using the derivatives of entropy with respect to $x$: $\frac{\partial S}{\partial x} = -f(x) \log f(x)$ and $\frac{\partial^2 S}{\partial x^2} = -(1+\log f(x))\frac{\partial f(x)}{\partial x} $ to obtain

\begin{equation}
\frac{\partial^2 S}{\partial x^2}\frac{\partial f(x)}{\partial t} = -\frac{1}{2N_e} \frac{\partial f(x)}{\partial x}\frac{\partial S}{\partial x}
\label{diffusion5}
\end{equation}

Now we come close to the conclusion given the approximation $S \approx 2x(1-x)$, we see that 
\begin{equation}
\frac{\partial^2 S}{\partial x^2} \approx -4
\end{equation}

to give

\begin{equation}
\frac{\partial f(x)}{\partial t} = \frac{1}{8N_e} \frac{\partial f(x)}{\partial x}\frac{\partial 2x(1-x)}{\partial x}
\label{diffusion6}
\end{equation}

and

\begin{equation}
\frac{\partial f(x)}{\partial t} = \frac{1}{4N_e} \frac{\partial f(x)}{\partial x}\frac{\partial x(1-x)}{\partial x}
\label{diffusion7}
\end{equation}

This is not the final diffusion equation. To complete this derivation we borrow from the approximate solution to $f(x,t)$ that Kimura derived in \cite{kimurastoch}

\begin{equation}
f(x,t) \approx 6p(1-p)e^{-t/2N}
\end{equation}

showing $f(x,t)$ depends only on $p$, the probability at $t=0$, and $t$ therefore we can determine that if $f(x)$ is not dependent on $x$.

\begin{equation}
\frac{\partial f(x)}{\partial t} = \frac{1}{4N_e} \frac{\partial}{\partial x}\frac{\partial f(x)x(1-x)}{\partial x}
\label{diffusion8}
\end{equation}

and finally

\begin{equation}
\frac{\partial f(x)}{\partial t} = \frac{1}{4N_e} \frac{\partial^2}{\partial x^2}f(x)x(1-x)
\label{diffusion9}
\end{equation}

\section{Channel Capacity of Genetic Information Inheritance in a Population}

This section is put into an appendix as an interesting mathematical excursion. Throughout this paper, I have striven to represent only the concepts most familiar to population genetics and which would prove most useful to theorists and practitioners whose main motivation is not just intellectual excursion but approaching real problems.

This section will be more speculative but in short, the use of information theory to reproduce population genetics opens avenues not only to more easily represent previously mathematically difficult concepts, but to couple previously unrelated ideas using information theory as a bridge. Claude Shannon defined channel capacity in terms of entropy in his landmark paper \cite{shannon}. The channel capacity is the maximum rate which a signal made up of symbols with a probability for each symbol can be transmitted. For a population, the channel can be represented as the combined effects of mating and offspring fitness with the next generation's allele frequency being indicative of the previous generations with evolutionary effects. There have been previous efforts to calculate the channel capacity of genetic information. Two investigations at the individual level genome, versus the population level in this paper, were done by Watkins \cite{channelcap1,channelcap2}. His investigation focuses on the channel capacity for the transmission of information at the level of the individual genome by selection (natural or artificial) given the genome length, allele distribution, and population size. Near $p=0.5$ he finds this channel capacity is directly proportional to the genome length $L$. He also shows that sexual reproduction allows a higher channel capacity than asexual reproduction.  A general idea of a channel capacity for evolution was also raised by science fiction writer Jonathan vos Post.

In terms of a source entropy, $H(X)$, a noiseless channel has a capacity $C$

\begin{equation}
C = NH(X)
\end{equation} 

where $N$ is the number of symbols per unit time. For a channel with noise, this capacity is equal to the entropy of the source minus the conditional entropy of the received signal $H(Y)$ termed $H_y(X)$

\begin{equation}
C = N(H(X) - H_y(X))
\end{equation} 

For a simple type of noise that changes a character with a fixed random probability, for example a bit flip from 1 to 0 or a mutation of an allele $p$ to $q$ or vice versa as seen in SNPs, the channel is the entropy of the source minus the conditional entropy. Note for all equations the natural logarithm is used but the channel capacity in bits/second can be derived by dividing the result by $\log 2$. For two alleles, $p$ and $q$ who have a probability $\mu$ of mutating into each other this channel capacity in a population of diploid organisms can be represented as

\begin{equation}
C = 2N_e(-p \log p - q \log q + \mu \log \mu + (1-\mu) \log (1-\mu))
\end{equation} 

or

\begin{equation}
C = 2N_e(S - S_m)
\end{equation} 

Using approximations, this equation can also be represented as

\begin{equation}
C = 2N_e(S - 2\mu) = 2N_e(h - 2\mu)
\end{equation} 

In other words, the channel capacity is represented by the entropy of the allele frequencies minus twice the mutation rate. So increasing the genetic diversity seems to increase channel capacity while mutation, which ironically also increases the genetic diversity, reduces it. Again, $S \approx h$ is only valid when $I=I'=0$.

Now we will calculate the channel capacity in special cases of population balance. For Hardy-Weinberg equilibrium obviously $C=2N_eS$  perpetually. In addition, with mutation the channel capacity is 0 for $S \approx h = 2\mu$. This should be expected for a signal-to-noise level of 0. Since in a completely homozygous population, mutation would introduce heterozygosity, an approximate signal-to-noise ratio can be hypothesized as

\begin{equation}
SNR = \frac{S}{S_m} - 1 = \frac{S}{2\mu} - 1 \approx \frac{pq}{\mu} - 1
\end{equation}

where the $pq/\mu$ approximation is only valid when there is no selection or non-random mating. For an $S$ several orders of magnitude larger than $\mu$ we can simply state

\begin{equation}
SNR = \frac{S}{2\mu} \approx \frac{pq}{\mu}
\end{equation}

$S/\mu$ may also be an acceptable approximation at these orders of magnitude.

Now for more detailed situations where entropy is known under balance conditions. For drift-mutation balance we have

\begin{equation}
C = 2N_e(h - 2\mu) = 2N_e(4N_e\mu  - 2\mu)
\end{equation} 

\begin{equation}
C = 4N_e\mu(2N_e - 1) \approx 8N_e^2\mu 
\end{equation} 

With drift and mutation balancing, we see several interesting effects. First, the channel capacity increases with the square of the effective population size, much faster than normal. More surprising though is the channel capacity becomes directly proportional to the mutation rate so in this case, increasing the rate of mutation actually increases the channel capacity. This is to be expected since in drift-mutation balance, mutation is the only force maintaining variation which drift would otherwise push to 0 over time. Migration-drift balance, using 
\begin{equation}
h=\frac{2N_em}{1+2N_em}h^*
\end{equation}

gives

\begin{equation}
C = 4N_e\bigg[\frac{N_emh^*-\mu(1+2N_eM)}{1+2N_em}\bigg]
\end{equation}

For the case of selection-mutation balance, one can show given $I'=S_{t-1} - S_t(p|q)=2\mu$ the channel capacity is proportional to the conditional entropy

\begin{equation}
C = 2N_eS(p|q)
\label{channelselect}
\end{equation}

This is an especially interesting result. Under all conditions, $S(p|q) \leq S$ which demonstrates that natural selection acts on the channel capacity in an equivalent manner to a filter by which stronger selection (lower conditional entropy) reduces the channel capacity acting as a filter on the amount of variation which can propagate between generations in a population discarding $S - S(p|q)$ variation. Stronger effects of natural selection when balanced with mutation, cancel out the mutation effects so the channel capacity depends only on the conditional entropy and the induced effects of selection (or nonrandom mating). In fact, in all of these examples, the channel capacity divided by the number of alleles ($2N_e$) is the maximum entropy  the population allele frequencies can maintain between generations without change. By definition, any source entropy rate above the channel capacity can not be transmitted without error and in this case, the allele frequencies would be forced to change back towards the entropy representing the channel capacity. For drift-mutation balance this is $S=4N_e\mu$ and for selection-mutation balance this is $S=S(p|q)$. 

Finally, we have the balance for combined effects of all forces which gives

\begin{equation}
C = 2N_e\bigg[\frac{2\mu + S_t(p|q) + mS^*}{1+\frac{1}{2N_e} + m}-2\mu \bigg]
\end{equation}

reducing to equation \ref{channelselect} for large populations with no gene flow or drift.

\section{Maximum Entropy and Hardy-Weinberg Equilibrium}
Since many researchers may not have access to or be able to translate the paper by Wang et. al. \cite{HWMaxEnt} that derives HWE from the maximum joint entropy of an allele distribution, a short derivation is included below.

The quantity to be maximized is $S_2$

\begin{equation}
S_2 = -\sum_{i=1}^n \sum_{j=1}^n P(i,j)\log P(i,j)
\end{equation}

subject to the constraints

\begin{equation}
\sum_{i=1}^n \sum_{j=1}^n P(i,j) = 1
\end{equation}

\begin{equation}
\frac{1}{2}\sum_{j=1}^n (P(i,j) + P(j, i)) = P(i)
\end{equation}

Using the method of Lagrange multipliers, we can create a Lagrange function, 
\begin{eqnarray}
G(p,q) &=& -\sum_{i=1}^n \sum_{j=1}^n P(i,j)\log P(i,j) + (\ln\lambda_0 + 1)\Bigg[\sum_{i=1}^n \sum_{j=1}^n P(i,j) - 1\Bigg] \nonumber \\ 
&+& \sum^n_{i=1} \ln \lambda_i \Bigg[\frac{1}{2}\sum_{j=1}^n (P(i,j) + P(j, i)) - P(i)\Bigg] \nonumber\\
\end{eqnarray}

Taking $\frac{\partial G}{\partial P(i,j)}=0$ we get

\begin{equation}
\ln P(i,j) - \ln \lambda_0 - \frac{1}{2}(\ln \lambda_i + \lambda_j) = 0
\end{equation}

which solves to

\begin{equation}
P(i,j) = \lambda_0 \sqrt{\lambda_i\lambda_j}
\end{equation}

Using the indexes we can easily see that $P(1,1) = \lambda_0\lambda_1$, $P(1,2)=P(2,1)=\lambda_0\sqrt{\lambda_1\lambda_2}$, and $P(2,2)=\lambda_0\lambda_2$.

Our constraint equations can thus be restated as follows:
\begin{eqnarray}
\lambda_0(\lambda_1+\lambda_2+2\sqrt{\lambda_1\lambda_2})=1 \nonumber\\
\lambda_0(\lambda_1+\sqrt{\lambda_1\lambda_2}) = P(1)\nonumber\\
\lambda_0(\lambda_2+\sqrt{\lambda_1\lambda_2}) = P(2)\nonumber\\
\end{eqnarray}

Doing the math on $P(1)$ and $P(2)$ you can clearly see that $P(1,1) = \lambda_0\lambda_1 = P(1)^2$, $P(2,2) = \lambda_0\lambda_2 = P(2)^2$, and $P(1,2) = P(2,1) = \lambda_0\sqrt{\lambda_1{\lambda_2}} = P(1)P(2)$
\end{document}